\documentclass[preprint,12pt]{imsart}

\usepackage[dvipdfmx]{graphicx}
\usepackage{amssymb,amsmath,bm}
\usepackage{amsthm}
\usepackage{subfigure}
\usepackage{enumerate}
\usepackage{multirow}
\usepackage{threeparttable}

\usepackage{natbib}
\def\cite{\citet}

\usepackage[dvips]{geometry}
\startlocaldefs
\newlength{\figwidth}
\newcommand{\myfig}[1]{\includegraphics[width=\figwidth,clip]{#1}} %% with figures
\newtheorem{theo}{Theorem}
\newtheorem{lemm}{Lemma}
\newcommand{\argmin}{{\rm arg}\mathop{\rm min\,}\limits}
\newcommand{\argmax}{{\rm arg}\mathop{\rm max\,}\limits}
\newcommand{\trace}{\mathop{\rm tr}\nolimits}
\newcommand{\bmh}[1]{\bm{\hat {#1}}}
\newcommand{\bmb}[1]{\bm{\bar {#1}}}
\newcommand{\bmt}[1]{\bm{\tilde {#1}}}
\endlocaldefs

\begin{document}

%%% title and author
\begin{frontmatter}
\title{An information criterion for model selection with missing data via complete-data divergence}
\runtitle{An information criterion for incomplete data}
\author[add1,add2]{\fnms{Hidetoshi}
\snm{Shimodaira}\corref{}\ead[label=e1]{shimo@sigmath.es.osaka-u.ac.jp}\thanksref{t1}}
\and
\author[add1]{\fnms{Haruyoshi} \snm{Maeda}\thanksref{t2}} 
\thankstext{t1}{Supported in part by JSPS KAKENHI Grant (24300106, 16H02789).}
\thankstext{t2}{Currently at Kawasaki Heavy Industries, Ltd.
 1-1, Kawasaki-cho, Akashi City, 673-8666, Japan.}
\address[add1]{Division of Mathematical Science, Graduate School of Engineering Science,
Osaka University\\
1-3 Machikaneyama-cho, Toyonaka, Osaka, Japan\\
\printead{e1}
\address[add2]{RIKEN Center for Advanced Integrated Intelligence Research\\
1-4-1 Nihonbashi, Chuo-ku Tokyo, Japan}
}

% \author{\fnms{John} \snm{Smith}\corref{}\ead[label=e1]{smith@foo.com}\thanksref{t1}}
% \thankstext{t1}{Thanks to somebody} 
% \address{line 1\\ line 2\\ printead{e1}}
% \affiliation{Some University}

\runauthor{H.~SHIMODAIRA AND H.~MAEDA}

\begin{abstract}

We derive an information criterion to select a parametric model of complete-data distribution
when only incomplete or partially observed data is available. Compared with AIC, our new criterion
has an additional penalty term for missing data, which is expressed by the Fisher information matrices of
complete data and incomplete data. We prove that our criterion is an asymptotically unbiased
estimator of complete-data divergence, namely, the expected Kullback-Leibler divergence between
the true distribution and the estimated distribution for complete data, whereas AIC is that for the
incomplete data. Information criteria PDIO (Shimodaira 1994) and AICcd (Cavanaugh and Shumway 1998)
have been previously proposed to estimate complete-data divergence, and they
have the same penalty term. The additional penalty term of our criterion for missing data turns
out to be only half the value of that in PDIO and AICcd. The difference in the penalty term
is attributed to the fact that our criterion is derived under a weaker assumption.
A simulation study with the weaker assumption shows that
our criterion is unbiased while the other two criteria are biased. In addition,
we review the geometrical view of alternating minimizations of the EM algorithm.
This geometrical view plays an important role in deriving our new criterion.

\end{abstract}


\end{frontmatter}

\section{Introduction} \label{sec:intro}

Modeling complete data $X=(Y,Z)$ is often preferable to modeling incomplete or
partially observed data $Y$ when missing data $Z$ is not observed. The
expectation-maximization (EM) algorithm \citep{dempster1977maximum} computes the  maximum likelihood
estimate of parameter vector $\bm{\theta}$ for a parametric model of the probability distribution of $X$. In this research, we
consider the problem of model selection in such situations. For mathematical simplicity, 
we assume that $X$ consists of independent and identically distributed random vectors. More specifically, $X
= (\bm{x}_1,\bm{x}_2,\ldots,\bm{x}_n)$, and the complete-data distribution is modeled as
$\bm{x}_1,\bm{x}_2,\ldots,\bm{x}_n \sim p_x(\bm{x}; \bm{\theta})$. Each vector is decomposed as
$\bm{x}^T = (\bm{y}^T,\bm{z}^T)$, and the marginal distribution is expressed as $p_y(\bm{y};
\bm{\theta}) = \int p_x(\bm{y}, \bm{z} ; \bm{\theta})\,d\bm{z}$, where 
$T$ denotes the matrix transpose and the integration is over all possible values of $\bm{z}$. We
formally treat $\bm{y}, \bm{z}$ as continuous random variables with the joint density function $p_x$.
However, when they are discrete random variables, 
the integration should be replaced with a summation of the probability functions.
We use symbols such as $p_x$ and $p_y$ for both the continuous and discrete cases, and simply refer to them as distributions.

The log-likelihood function is $\ell_y(\bm{\theta}) = \sum_{t=1}^n \log p_y(\bm{y}_t ;
\bm{\theta})$ with the parameter vector $\bm{\theta}=(\theta_1,\ldots,\theta_d)^T\in\mathbb{R}^d$.
We assume that the model is identifiable, and the parameter is restricted to $ \Theta \subset \mathbb{R}^d$.
Then the maximum likelihood estimator (MLE) of $\bm{\theta}$ is defined by $\bmh{\theta}_y = \argmax_{\bm{\theta}\in\Theta}
\ell_y(\bm{\theta})$. The dependence of $\ell_y(\bm{\theta})$ and $\bmh{\theta}_y$ on
$Y=(\bm{y}_1,\ldots,\bm{y}_n)$ is suppressed in the notation. \cite{akaike1974new} proposed the
information criterion
\[
  {\rm AIC} = -2 \ell_y(\bmh{\theta}_y ) + 2 d
\]
for model selection. The first term measures the goodness of fit, whereas the second term is interpreted
as a penalty for model complexity. The AIC values for candidate models are computed, and then the model that minimizes AIC is selected. This information criterion estimates the expected discrepancy between
the unknown true distribution of $\bm{y}$, which is denoted as $q_y$, and the estimated distribution $p_y(\bmh{\theta}_y)$.
This discrepancy is measured by the incomplete-data Kullback-Leibler divergence.

In this study, we work on the complete-data Kullback-Leibler divergence instead of the incomplete-data
counterpart. An information criterion to estimate the expected discrepancy
between the unknown true distribution of $\bm{x}$, which is denoted as $q_x$, and the estimated
distribution $p_x(\bmh{\theta}_y)$ is derived. This approach makes sense when modeling complete data more precisely describes the part being examined.
Similar attempts are found
in the literature.
\cite{shimodaira1994new} proposed the information criterion PDIO (predictive divergence for
incomplete observation models)
\[
   {\rm PDIO} = -2 \ell_y(\bmh{\theta}_y ) + 2 \trace(I_x(\bmh{\theta}_y) I_y(\bmh{\theta}_y)^{-1}).
\]
The two matrices in the penalty term are the Fisher information matrices for complete data and
incomplete data. They are defined by
\begin{gather*}
  I_x(\bm{\theta}) = - \int p_x(\bm{x} ; \bm{\theta}) 
    \frac{\partial^2 \log p_x(\bm{x} ; \bm{\theta})}
    {\partial \bm{\theta} \partial \bm{\theta}^T}  \,d\bm{x}, \\
  I_y(\bm{\theta}) = - \int p_y(\bm{y} ; \bm{\theta})
     \frac{\partial^2 \log p_y(\bm{y} ; \bm{\theta})}
     {\partial \bm{\theta} \partial \bm{\theta}^T}  \,d\bm{y}.
\end{gather*}
Let $p_{z|y}(\bm{z}| \bm{y} ; \bm{\theta}) = p_x(\bm{y},\bm{z} ; \bm{\theta})/p_y(\bm{y} ; \bm{\theta})$ be the conditional distribution of $\bm{z}$ given $\bm{y}$, and $I_{z|y}(\bm{\theta}) = I_x(\bm{\theta} )- I_y(\bm{\theta})$ be the Fisher information matrix for $p_{z|y}$. Since $I_{z|y}(\bm{\theta})$ is nonnegative definite, we have
$\trace( I_x(\bm{\theta}) I_y(\bm{\theta})^{-1}  )  = \trace( (I_y(\bm{\theta}) +
I_{z|y}(\bm{\theta}  ) ) I_y(\bm{\theta})^{-1}  )  =  d + \trace(  I_{z|y}(\bm{\theta} )
I_y(\bm{\theta})^{-1}  )  \ge d$. Thus, the nonnegative difference 
\[
  {\rm PDIO} - {\rm AIC} = 2 \trace( I_{z|y}(\bmh{\theta}_y ) I_y(\bmh{\theta}_y)^{-1}  ) 
\]
is interpreted as the additional penalty for
missing data. There are similar attempts in the literature \citep{cavanaugh1998akaike,seghouane2005criterion,claeskens2008variable,yamazaki2014asymptotic}.
In particular, \cite{cavanaugh1998akaike} proposed another information criterion
\[
  {\rm AIC}_{cd} = -2 Q(\bmh{\theta}_y; \bmh{\theta}_y)
  + 2 \trace(I_x(\bmh{\theta}_y) I_y(\bmh{\theta}_y)^{-1})
\]
by replacing $\ell_y(\bmh{\theta}_y )$ in PDIO with $Q(\bmh{\theta}_y; \bmh{\theta}_y)$ to measure
the goodness of fit. It should be noted that cd stands for complete data. This is the
function introduced in \cite{dempster1977maximum} for the EM algorithm, and is defined by
\[
  Q(\bm{\theta}_2; \bm{\theta}_1) =  \sum_{t=1}^n 
  \int p_{z|y} (\bm{z}| \bm{y}_t ; \bm{\theta}_1)
  \log p_x(\bm{y}_t, \bm{z} ; \bm{\theta}_2)\,d\bm{z}.
\]

We recently found that the assumption in \cite{shimodaira1994new} to derive PDIO is unnecessarily strong. Additionally, the same assumption explains the derivation of ${\rm AIC}_{cd}$.
In this paper, we derive a new information criterion 
 under a weaker assumption. The updated version of PDIO is
\[
   {\rm AIC}_{x;y} = -2 \ell_y(\bmh{\theta}_y ) + d +  
   \trace(I_x(\bmh{\theta}_y) I_y(\bmh{\theta}_y)^{-1}).
\]
The first suffix $\bm{x}$ indicates that a random variable is used to measure the discrepancy, while the second suffix
$\bm{y}$ indicates a random variable is used for the observation. Then the additional penalty for missing data becomes
\begin{equation} \label{eq:aicxy-aic}
  {\rm AIC}_{x;y} - {\rm AIC} = \trace(  I_{z|y}(\bmh{\theta}_y ) I_y(\bmh{\theta}_y)^{-1}  ).
\end{equation}
The additional penalty is only half the value of that in PDIO.
In practice, the computation of ${\rm
AIC}_{x;y}$ as well as the related criteria PDIO and ${\rm AIC}_{cd}$ is not very difficult.
The SEM algorithm of \cite{meng1991using} provides a shortcut to compute the penalty term
$\trace(I_x(\bmh{\theta}_y) I_y(\bmh{\theta}_y)^{-1})$ without computing the two Fisher information
matrices as described in \cite{shimodaira1994new} and
\cite{cavanaugh1998akaike}.

To derive $ {\rm AIC}_{x;y} $, we first review the basic properties of Kullback-Leibler
divergence for incomplete data in Section~\ref{sec:projectiony}.
Section~\ref{sec:projectionx} considers those for complete data. Although these results are not new,
they are crucial for the argument in later sections. In particular, the geometrical view of alternating minimizations  \citep{csisz1984information,amari1995information} in
Section~\ref{sec:projectionx-dual} is important to understand why the goodness of fit term of
${\rm AIC}_{x;y}$ is expressed by the incomplete-data likelihood function instead of the
complete-data counterpart.

Section~\ref{sec:risk}, which begins the argument of model selection, discusses what the information criteria should estimate. In general, parametric models are misspecified, and we do
\emph{not} assume that the true distribution is expressed as $q_x = p_x(\bm{\theta}_0)$ using the
``true'' parameter value $\bm{\theta}_0$. However, the unbiasedness of ${\rm AIC}_{x;y}$ is
based on the assumption that $p_{z|y}(\bm{\theta})$ is correctly specified for $q_{z|y}$. In
Section~\ref{sec:information-criterion}, we derive our new information criterion. The argument is very
straightforward; it simply follows the argument for the robust version of AIC, which is also known as the Takeuchi
information criterion (TIC) that is described in \cite{burnham2002model} and \cite{konishi2008information}.
Section~\ref{sec:pdio-aiccd} compares the assumptions used to derive PDIO and ${\rm AIC}_{cd}$
to those of ${\rm AIC}_{x;y}$.
Section~\ref{sec:simulation} presents a simulation study to verify the theory.
Finally, Section~\ref{sec:conclusion} contains some concluding remarks. Proofs are deferred to the Appendix.

\section{Incomplete-data divergence} \label{sec:projectiony}

Here we review Kullback-Leibler divergence and the asymptotic distribution of MLE under
model misspecification \citep{white1982maximum}. 
Let $g_y$ and $f_y$ be the arbitrary probability
distributions of incomplete data.  The incomplete-data Kullback-Leibler divergence from $g_y$ to
$f_y$ is
\[  
   D_y(g_y ; f_y) =  - \int g_y(\bm{y}) (\log f_y(\bm{y}) - \log g_y(\bm{y}))\,d\bm{y},
\] 
where $D_y(g_y ; f_y)\ge0$ and the equality holds for $g_y = f_y$
\citep{csiszar1975divergence,amari2007methods}. 
The cross-entropy is
\[
  L_y(g_y ; f_y) = - \int g_y(\bm{y})  \log f_y(\bm{y})\,d\bm{y} 
\] 
and the entropy is $ L_y(g_y)  = L_y(g_y; g_y)$. Instead of minimizing $D_y(g_y ; f_y)= L_y(g_y ; f_y)
- L_y(g_y)$ with respect to $f_y$, we minimize $L_y(g_y ; f_y)$, because $L_y(g_y)$ is independent of $f_y$.

\begin{figure}[hbt] \begin{center}
  \myfig{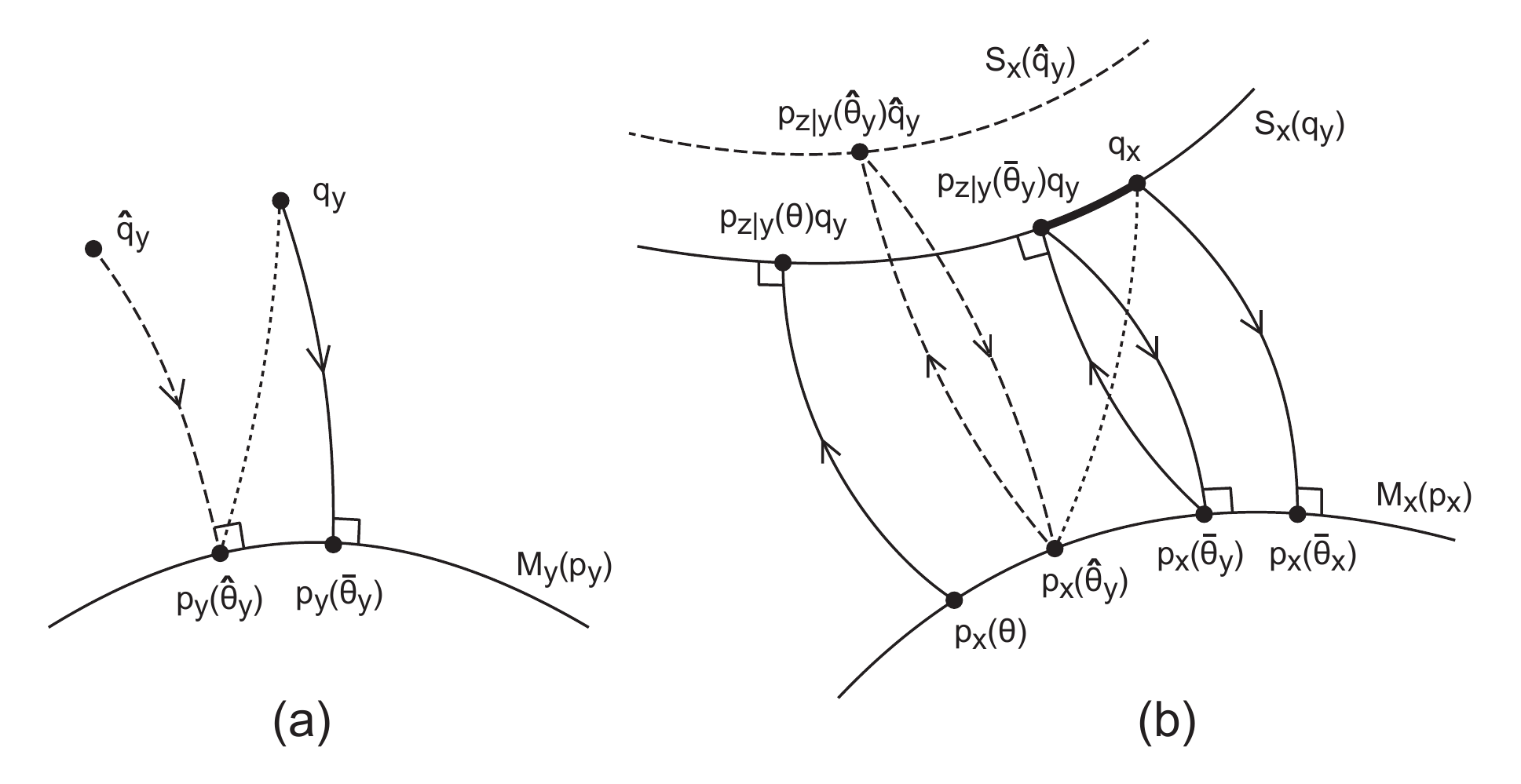}
  \caption{(a)~Space of incomplete-data probability distributions. Projection from $q_y$ to
  the model manifold $M_y(p_y)$ (arrow with a solid line), and that from $\hat q_y$ (arrow with a broken line) using eqs.~(\ref{eq:projectiony-to-model}) and
  (\ref{eq:mley}) in Section~\ref{sec:projectiony}, respectively. The dotted line indicates 
  $D_y(q_y ; p_y(\bmh{\theta}_y))$, which is the loss function for ${\rm risk}_{y;y} $.
  (b)~Space of complete-data probability
  distributions. Projection from $q_x$ to the model manifold $M_x(p_x)$ using
  eq.~(\ref{eq:projectionx-to-model}) in Section~\ref{sec:projectionx-to-model}. Projection from
  $p_x(\bm{\theta})$ to the data manifold $S_x(q_y)$ using eq.~(\ref{eq:projectionx-to-data}) in
  Section~\ref{sec:projectionx-to-data}.  Alternating projections between the two manifolds using
  eq.~(\ref{eq:projectionx-dual}) in Section~\ref{sec:projectionx-dual}. 
  The dotted line indicates $D_x(q_x ; p_x(\bmh{\theta}_y))$, which is the loss function for ${\rm risk}_{x;y} $.
  The bold segment indicates $ D_x(q_x ; p_{z|y}(\bmb{\theta}_y) q_y)$, which is assumed
  to be zero in (\ref{eq:asm-qzy}).
  }
  \label{fig:projections}  
\end{center} \end{figure}

For the true distribution $q_y$ and the parametric model $p_y(\bm{\theta})$,  we consider the
minimization of $D_y(q_y ; p_y(\bm{\theta}))$ with respect to $\bm{\theta}$. The optimal parameter
value is defined by
\[  
  \bmb{\theta}_y   = \argmin_{\bm{\theta}\in\Theta} L_y(q_y ; p_y(\bm{\theta})). 
\]
This minimization is interpreted geometrically as a ``projection'' of $q_y$ to the model 
manifold $M_y(p_y)$ as illustrated in Fig.~\ref{fig:projections} (a).  Let
$M_y(p_y) = \{p_y(\bm{\theta}) : \forall \bm{\theta}\in\Theta
\}$ be the set of $p_y(\bm{\theta})$ with all possible parameter values. Then the projection is
defined as
\begin{equation} \label{eq:projectiony-to-model} 
  \min_{f_y \in M_y(p_y)} D_y(q_y ; f_y) = D_y(q_y; p_y(\bmb{\theta}_y)). 
\end{equation} 
The projection $p_y(\bmb{\theta}_y)$ is the best approximation of $q_y$ in $M_y(p_y)$ when the discrepancy is
measured by the Kullback-Leibler divergence. We assume that the parametric model is
generally misspecified and $q_y \not\in  M_y(p_y)$. Later, we also consider the situation
where the parametric model is correctly specified and $q_y \in M_y(p_y)$. In the correctly specified case,
$\bmb{\theta}_y$ is the true parameter value in the sense that $q_y = p_y(\bmb{\theta}_y)$.

Similar to the optimal parameter value, the maximum likelihood estimator is interpreted as a projection of $\hat q_y$ to $M_y(p_y)$.
Let $ \hat q_y(\bm{y}) = \frac{1}{n} \sum_{t=1}^n \delta(\bm{y} - \bm{y}_t)$ be the empirical
distribution of $\bm{y}$ for the observed incomplete data $\bm{y}_1,\ldots, \bm{y}_n$. Here $\delta(\cdot)$
denotes the Dirac delta function for continuous random variables, or is simply the indicator function
for discrete random variables such that $\delta(\bm{y} - \bm{y}_t)=1$ for $\bm{y} = \bm{y}_t$  and
$\delta(\bm{y} - \bm{y}_t)=0$ otherwise. Then we can write $\ell_y(\bm{\theta}) = -n L_y(\hat q_y ;
p_y(\bm{\theta}))$. Thus,
\begin{equation} \label{eq:mley}
  \bmh{\theta}_y   = \argmin_{\bm{\theta}\in\Theta} L_y(\hat q_y ; p_y(\bm{\theta})). 
\end{equation}

We assume the regularity conditions of \cite{white1982maximum} for consistency 
and asymptotic normality of $\bmh{\theta}_y$ . More specifically, we assume all the
regularity conditions (A1) to (A6) for the true distribution $q_y$ and the model distribution
$p_y(\bm{\theta})$. In particular, $\bmb{\theta}_y$ is determined uniquely (i.e., identifiable) and
is interior to the parameter space $\Theta$. We assume that $I_y(\bm{\theta})$, $G_y(q_y;
\bm{\theta})$ and $H_y(q_y; \bm{\theta})$ defined below are nonsingular in the neighborhood of
$\bmb{\theta}_y$. Then \cite{white1982maximum} showed that, as $n\to\infty$ asymptotically,
$\bmh{\theta}_y  \stackrel{a.s.}{\to} \bmb{\theta}_y$ and
\begin{equation} \label{eq:mley-normal}
  \sqrt{n}\,(\bmh{\theta}_y - \bmb{\theta}_y ) \stackrel{d}{\to}  
  N(\bm{0}, H_y^{-1} G_y H_y^{-1}).
\end{equation}
The matrices are defined as $G_y = G_y(q_y; \bmb{\theta}_y)$ and $H_y = H_y(q_y;
\bmb{\theta}_y)$, where 
\begin{align*} 
  G_y(g_y;\bm{\theta}) &= \int g_y(\bm{y}) \frac{\partial
  \log p_y(\bm{y} ; \bm{\theta})}{\partial \bm{\theta} }     \frac{\partial \log p_y(\bm{y} ;
  \bm{\theta})}{\partial \bm{\theta}^T } \,d\bm{y},\\
  H_y(g_y;\bm{\theta}) &= - \int g_y(\bm{y})
  \frac{\partial^2 \log p_y(\bm{y} ; \bm{\theta})} 
   {\partial \bm{\theta} \partial \bm{\theta}^T} \,d\bm{y}.
\end{align*}
In the case of the correct specification $q_y = p_y(\bmb{\theta}_y)$, the matrices become $G_y=H_y=I_y(\bmb{\theta}_y)$.

\section{Complete-data divergence} \label{sec:projectionx}

Here we review Kullback-Leibler divergence for complete data when only incomplete data
can be observed \citep{csisz1984information,amari1995information}. 

\subsection{Projection to the model manifold} \label{sec:projectionx-to-model}

Let $g_x$ and $f_x$ be the arbitrary probability distributions of complete data.  The complete-data
Kullback-Leibler divergence from $g_x$ to $f_x$ is  
\[
  D_x(g_x ; f_x) =  - \int g_x(\bm{x}) (\log f_x(\bm{x}) - \log g_x(\bm{x}))\,d\bm{x}. 
\] 
All the arguments of incomplete data in Section~\ref{sec:projectiony} apply to complete data by
replacing $\bm{y}$ with $\bm{x}$ in the notation. For example, we write $ D_x(g_x ; f_x) = L_x(g_x ; f_x) -
L_x(g_x)$ with $ L_x(g_x ; f_x) = - \int g_x(\bm{x})  \log f_x(\bm{x})\,d\bm{x}$ and $L_x(g_x)  =
L_x(g_x; g_x) $. The projection of $q_x$ to the model manifold $M_x(p_x) = \{p_x(\bm{\theta}) :
\forall \bm{\theta} \in \Theta \}$ is defined as
\begin{equation} \label{eq:projectionx-to-model} 
  \min_{f_x \in M_x(p_x)} D_x(q_x ; f_x) = D_x(q_x; p_x(\bmb{\theta}_x)) 
\end{equation}
with $ \bmb{\theta}_x   = \argmin_{\bm{\theta}} L_x(q_x ; p_x(\bm{\theta}))$.
Figure~\ref{fig:projections} (b) shows a geometric illustration. Note that 
$\bmb{\theta}_x \neq \bmb{\theta}_y$ and $p_x(\bmb{\theta}_x ) \neq p_x(\bmb{\theta}_y)$ in general.

\subsection{Projection to the data manifold} \label{sec:projectionx-to-data}

The following simple lemma helps understand how the incomplete-data
divergence and the complete-data divergence are related.
\begin{lemm} \label{lem:dx-decomp} 
For two distributions $g_x(\bm{x})$ and $f_x(\bm{x})$, we have
\begin{equation} \label{eq:dx-decomp1} 
  D_x(g_x ; f_x) = D_x(g_x ; f_{z|y} g_y) + D_y(g_y ; f_y), 
\end{equation} 
where $f_{z|y} g_y$ represents the distribution $f_{z|y}(\bm{z}|\bm{y}) g_y(\bm{y})$.  Therefore,
the difference of the two divergences is $ D_x(g_x ; f_x) - D_y(g_y ; f_y)  = D_x(g_x ; f_{z|y}
g_y)$, which is zero if $g_{z|y}=f_{z|y}$. For an arbitrary distribution $h_x(\bm{x})$, the last term
in (\ref{eq:dx-decomp1}) is expressed as
\begin{equation} \label{eq:dy-dx} 
  D_y(g_y ; f_y) = D_x(h_{z|y} g_y ; h_{z|y} f_y). 
\end{equation}
In particular, choosing $h_x = f_x$ gives $D_y(g_y ; f_y) = D_x(f_{z|y} g_y ; f_x)$, and
\begin{equation} \label{eq:dx-decomp2} 
  D_x(g_x ; f_x) = D_x(g_x ; f_{z|y} g_y) + D_x(f_{z|y} g_y ; f_x). 
\end{equation}
\end{lemm}

We consider the set of all probability distributions $g_x$ with the same marginal distribution
$g_y = q_y$ for a specified $q_y$. This set is denoted as $S_x(q_y) = \{ g_{z|y} q_y : \forall g_{z|y}\}$. Note that the elements of
$S_x(q_y)$ are written as $g_{z|y} q_y$ with arbitrary $g_{z|y}$ because $\int
g_{z|y}(\bm{z}|\bm{y}) q_y(\bm{y}) \,d\bm{z} = q_y(\bm{y})$. Equations~(88) and (57) in
\cite{amari1995information} are $S_x(\hat q_y)$ and its restriction to a finite dimensional model, respectively,
and are called the observed data (sub)manifold there. 
Here, we call $S_x(q_y)$ the expected data manifold and $S_x(\hat q_y)$ the observed data manifold,
although it may be abuse of the word ``manifold'' for subsets with infinite dimensions.

The projection of $p_x(\bm{\theta})$ to $S_x(q_y)$ should be defined to minimize the complete-data
divergence over $S_x(q_y)$, but the roles of $g_x$ and $f_x$ in $D_x(g_x;f_x)$ are exchanged from
those of (\ref{eq:projectionx-to-model}). We minimize $D_x(g_x; p_x(\bm{\theta}))$ over $g_x\in
S_x(q_y)$. By letting $g_x \in S_x(q_y)$ and $f_x = p_x(\bm{\theta})$ in (\ref{eq:dx-decomp1}),
\[
D_x(g_x; p_x(\bm{\theta})) = D_x(g_{z|y} q_y; p_{z|y} (\bm{\theta}) q_y) + D_y(q_y; p_y(\bm{\theta})),
\]
which is minimized when $g_{z|y} = p_{z|y}(\bm{\theta}) $. Therefore, the projection gives the
minimum value as
\begin{equation} \label{eq:projectionx-to-data}
    \min_{g_x \in S_x(q_y)} D_x(g_x ; p_x(\bm{\theta})) =  D_y(q_y ; p_y(\bm{\theta})).
\end{equation}
Using (\ref{eq:dx-decomp2}), the minimum value can also be written as
$  D_y(q_y ; p_y(\bm{\theta})) = D_x(p_{z|y}(\bm{\theta}) q_y ; p_x(\bm{\theta}))$.

\subsection{Alternating projections between the two manifolds} \label{sec:projectionx-dual}

The optimal parameter $\bmb{\theta}_y$ of the incomplete data is interpreted as a dual or alternate
minimization problem of complete-data divergence. By minimizing (\ref{eq:projectionx-to-data})
over $\bm{\theta}\in\Theta$, we define the alternating projections between $S_x(q_y)$ and $M_x(p_x)$
as
\begin{equation} \label{eq:projectionx-dual}
   \min_{f_x \in M_x(p_x)} \min_{g_x \in S_x(q_y)} D_x(g_x ; f_x) = D_y(q_y ; p_y(\bmb{\theta}_y)),
\end{equation}
where the minimum is attained by $g_x = p_{z|y}(\bmb{\theta}_y) q_y$ and $f_x =
p_x(\bmb{\theta}_y)$. See eq.~(65) in \cite{amari1995information}. This implies that
$p_{z|y}(\bmb{\theta}_y) q_y$ is the best approximation of $q_x$ when the two manifolds
$S_x(q_y)$ and $M_x(p_x)$ are known, while $p_x(\bmb{\theta}_y)$ is the best approximation of $q_x$ in
$M_x(p_x)$. This interpretation is the key to understanding our problem.

The above mentioned geometrical interpretation corresponds to the well known fact that the EM
algorithm of \cite{dempster1977maximum} is alternating projections between $S_x(\hat q_y)$ and
$M_x(p_x)$. See \cite{csisz1984information},
\cite{byrne1992alternating}, \cite{amari1995information}, and
\cite{ip2000note}. Starting from the initial value $\bm{\theta}^{(1)}$, the EM algorithm computes a
sequence of the parameter values $\{\bm{\theta}^{(s)}; s=1,2,\ldots \}$ by the updating formula $\bm{\theta}^{(s+1)} =
\argmax_{\bm{\theta}\in\Theta} Q(\bm{\theta};   \bm{\theta}^{(s)} )$.  It follows from $
L_x(p_{z|y}(\bm{\theta}_1) \hat q_y;  p_x(\bm{\theta}_2))  = -Q(\bm{\theta}_2 ; \bm{\theta}_1) / n $
that
\[
  \bm{\theta}^{(s+1)}  =
\argmin_{\bm{\theta}\in\Theta} L_x(p_{z|y}(\bm{\theta}^{(s)}) \hat q_y ;  p_x(\bm{\theta})  ),
\]
meaning $p_x(\bm{\theta}^{(s+1)})$ is the projection from $p_{z|y}(\bm{\theta}^{(s)}) \hat q_y$ to
$M_x(p_x)$. Alternatively, $p_{z|y}(\bm{\theta}^{(s)}) \hat q_y$  is the projection from
$p_x(\bm{\theta}^{(s)})$ to $S_x(\hat q_y)$. 
Thus, the converging point of the alternating projections satisfies
\begin{equation} \label{eq:mley-em}
  \bmh{\theta}_y  =
\argmin_{\bm{\theta}\in\Theta} L_x(p_{z|y}(\bmh{\theta}_y) \hat q_y ;  p_x(\bm{\theta})  ).
\end{equation}

\section{Risk functions for model selection} \label{sec:risk}

By looking at the incomplete-data distributions, the discrepancy between the true distribution $q_y$
and our estimation $p_y(\bmh{\theta}_y)$ is measured by the incomplete-data divergence $D_y(q_y ;
p_y(\bmh{\theta}_y))$. If we take it as the loss function, the expected loss-function, or the risk
function, will measure the discrepancy in the long run.  Then AIC and its variants are derived as
estimators of
\begin{equation} \label{eq:truerisky}
  {\rm risk}_{y;y}  =  E \{  D_y(q_y ; p_y(\bmh{\theta}_y)  )  \}.
\end{equation}
The expectation is evaluated with respect to $q_x$, although it involves only $q_y$ here.
This is the standard approach in the
literature \citep{akaike1974new,bozdogan1987model,burnham2002model,konishi2008information}.

\cite{shimodaira1994new} and \cite{cavanaugh1998akaike} proposed another approach, which employs
the complete-data divergence $D_x(q_x; p_x(\bmh{\theta}_y))$ to measure the discrepancy between
the complete-data distributions $q_x$  and $p_x(\bmh{\theta}_y)$. Using the complete-data divergence as the loss function, the risk function becomes
\begin{equation} \label{eq:trueriskx}
  {\rm risk}_{x;y}  =  E \{  D_x(q_x ; p_x(\bmh{\theta}_y)  )  \}.
\end{equation}
The first suffix $\bm{x}$ indicates the random
variable for the loss function, while the second suffix $\bm{y}$ indicates the random variable for
the observation.

However, estimating (\ref{eq:trueriskx}) is difficult. The complete-data empirical
distribution $\hat q_x(\bm{x}) =  \frac{1}{n} \sum_{t=1}^n \delta(\bm{x} - \bm{x}_t)$ is unknown;
we only know that $\hat q_x$ is somewhere in the observed data manifold
$S_x(\hat q_y)$. Considering the limiting situation of $n\to\infty$, we may only know that the true
distribution is somewhere in the expected data manifold: $q_x \in S_x(q_y)$. Then  the best
substitute for $q_x$ is 
\begin{equation} \label{eq:asm-qx}
  q_x = p_{z|y}(\bmb{\theta}_y) q_y
\end{equation}
as suggested by (\ref{eq:projectionx-dual})
from the viewpoint of the alternating projections in Section~\ref{sec:projectionx-dual}. 
To estimate (\ref{eq:trueriskx}), we assume that (\ref{eq:asm-qx}) holds in this paper.
This assumption is rephrased as 
\[
  D_x(q_x ; p_{z|y}(\bmb{\theta}_y) q_y) = 0
\]
or equivalently
\begin{equation} \label{eq:asm-qzy}
 q_{z|y} =  p_{z|y}(\bmb{\theta}_y),
 \end{equation}
implying that $p_{z|y}(\bm{\theta})$ is correctly specified for
$q_{z|y}$ and that $\bmb{\theta}_x = \bmb{\theta}_y$, because the two projections from $q_x$ and
$p_{z|y}(\bmb{\theta}_y)q_y$ to $M_x(p_x)$ become identical as illustrated in Fig.~\ref{fig:projections} (b). 
Because it is impossible to know how much $q_{z|y}$ actually deviates from $p_{z|y}(\bmb{\theta}_y)$ when
$Z=(\bm{z}_1,\ldots,\bm{z}_n)$ is missing \emph{completely}, we assume
(\ref{eq:asm-qzy}) in the following argument to derive ${\rm AIC}_{x;y}$. 
Note that assumption (\ref{eq:asm-qzy}) holds with $\bmb{\theta}_x = \bmb{\theta}_y=\bm{\theta}_0$ in the case of the correct specification where $q_x = p_x(\bm{\theta}_0)$.

We are now ready to derive  ${\rm AIC}_{x;y}$ as an estimator of $2n\,{\rm risk}_{x;y} $. The
arguments in Lemma~\ref{lem:riskxy} and Theorem~\ref{thm:riskxy-hat} almost duplicate that used
to derive TIC mentioned in \cite{burnham2002model} and \cite{konishi2008information}. However, it
should be noted that  in Lemma~\ref{lem:riskxy}  the first term of ${\rm risk}_{x;y} $  is expressed by
the incomplete-data divergence instead of the complete-data divergence. A point for proving the
lemma is that
\begin{equation}  \label{eq:dxisdy}
  D_x(q_x; p_x(\bmb{\theta}_y))= D_x(q_x ; p_{z|y}(\bmb{\theta}_y) q_y) + D_y(q_y ;
  p_y(\bmb{\theta}_y)  )  =  D_y(q_y ; p_y(\bmb{\theta}_y)  ),
\end{equation}
which follows from Lemma~\ref{lem:dx-decomp} and the assumption~(\ref{eq:asm-qzy}).
$D_x(q_x; p_x(\bmb{\theta}_y))$ on the left-hand side 
is the amount of misspecification of $p_x(\bm{\theta})$, and can be
decomposed into the two parts: $D_x(q_x ; p_{z|y}(\bmb{\theta}_y) q_y)$ and $D_y(q_y ; p_y(\bmb{\theta}_y)  )$, which are
the contribution of $p_{z|y}(\bm{\theta})$ and $p_y(\bm{\theta})$, respectively.
To estimate (\ref{eq:trueriskx}), instead of estimating $D_x(q_x ; p_{z|y}(\bmb{\theta}_y) q_y)$, we ignore it.

\begin{lemm} \label{lem:riskxy}
Assume the regularity conditions of \cite{white1982maximum} mentioned in
Section~\ref{sec:projectiony}, and also assume that (\ref{eq:asm-qzy}) holds.
Then the expected loss is asymptotically expanded as
\begin{equation} \label{eq:riskxy-expand}
  {\rm risk}_{x;y} = D_y(q_y ; p_y(\bmb{\theta}_y)  ) 
    +\frac{1}{2n} \trace(H_x H_y^{-1} G_y H_y^{-1}) + O(n^{-3/2}).
\end{equation}
The matrices $G_y$ and $H_y$ are those defined in Section~\ref{sec:projectiony}, and
$H_x = H_x(p_{z|y}(\bmb{\theta}_y)q_y; \bmb{\theta}_y)$ with
\[
  H_x(g_x; \bm{\theta}) = - \int g_x(\bm{x}) \frac{\partial^2 \log p_x(\bm{x} ; \bm{\theta})}
                                                  {\partial \bm{\theta} \partial \bm{\theta}^T}
     \,d\bm{x}.
\]
The dominant term in (\ref{eq:riskxy-expand}) is also expressed as
$D_y(q_y ; p_y(\bmb{\theta}_y)  )  = L_y(q_y ; p_y(\bmb{\theta}_y)  ) - L_y(q_y ) $
using the cross-entropy.
\end{lemm}

\section{Information criteria} \label{sec:information-criterion}

Let us define an information criterion as an estimator of ${\rm risk}_{x;y}$.
\begin{equation} \label{eq:hat-riskxy}
     \widehat{\rm risk}_{x;y} = L_y(\hat q_y ; p_y(\bmh{\theta}_y)  ) - L_y(q_y ) 
    +\frac{1}{2n} \trace(G_y H_y^{-1})
    +\frac{1}{2n} \trace(H_x H_y^{-1} G_y H_y^{-1}),
\end{equation}
where the matrices $G_y$, $H_y$ and $H_x$ may be replaced by their consistent estimators with error
$O_p(n^{-1/2})$. When $\bm{x}\equiv \bm{y}$, (\ref{eq:hat-riskxy}) reduces to
\begin{equation} \label{eq:hat-riskyy}
  \widehat{\rm risk}_{y;y} = L_y(\hat q_y ; p_y(\bmh{\theta}_y)  ) - L_y(q_y) 
    +\frac{1}{n} \trace(G_y H_y^{-1}),
\end{equation}
which corresponds to the Takeuchi information criterion (TIC) for estimating ${\rm risk}_{y;y}$
mentioned in \cite{burnham2002model} and \cite{konishi2008information}. In model selection,
we ignore $L_y(q_y)$, because all candidate models have the same
value. The first term $ L_y(\hat q_y ; p_y(\bmh{\theta}_y)  )  = -\ell_y(\bmh{\theta}_y)/n$ of order
$O_p(1)$ measures the goodness of fit, while the last two terms of order $O(n^{-1})$ are interpreted as the
penalty of model complexity. Our estimator is justified by the following theorem.
\begin{theo} \label{thm:riskxy-hat}
Assume the regularity conditions of \cite{white1982maximum} mentioned in
Section~\ref{sec:projectiony}, and also assume that (\ref{eq:asm-qzy}) holds. Then
we have 
\begin{equation} \label{eq:lyy-expand}
   L_y(q_y ; p_y(\bmb{\theta}_y)) 
   = E\{  L_y(\hat q_y ; p_y(\bmh{\theta}_y)) \} + \frac{1}{2n}\trace( G_y H_y^{-1}) + O(n^{-3/2}),
\end{equation}
and therefore
\begin{equation} \label{eq:riskxy-unbiased}
   E\{ \widehat{\rm risk}_{x;y}  \} =  {\rm risk}_{x;y}  + O(n^{-3/2}).
\end{equation}
Thus, the estimator is unbiased asymptotically up to terms of order $O(n^{-1})$.
\end{theo}

In the case of the correct specification where $q_y = p_y(\bmb{\theta}_y)$ for the incomplete-data distribution,
we have $G_y=H_y=I_y(\bmb{\theta}_y)$, and the information matrix is consistently estimated by $I_y(\bmh{\theta}_y)$. Assuming
(\ref{eq:asm-qzy}), this implies that $q_x = p_x(\bmb{\theta}_x)$ is correctly specified for the complete-data
distribution. Hence, $H_x=I_x(\bmb{\theta}_y)$ is consistently estimated by $I_x(\bmh{\theta}_y)$.
For model selection, we assume that $p_y(\bm{\theta})$ is misspecified for $q_y$ in general.
However, these equations may approximately hold if $p_y(\bmb{\theta}_y)$ is a good approximation of
$q_y$. By substituting $G_y \approx H_y \approx I_y(\bmb{\theta}_y)$ and $H_x \approx
I_x(\bmb{\theta}_y)$ into (\ref{eq:hat-riskxy}) and (\ref{eq:hat-riskyy}), we have
\[
    \widehat{\rm risk}_{x;y} \approx L_y(\hat q_y ; p_y(\bmh{\theta}_y)  ) - L_y(q_y ) 
    +\frac{d}{2n}
    +\frac{1}{2n} \trace(I_x(\bmh{\theta}_y) I_y(\bmh{\theta}_y)^{-1}),
\]
and
\[
    \widehat{\rm risk}_{y;y} \approx L_y(\hat q_y ; p_y(\bmh{\theta}_y)  ) - L_y(q_y ) 
    +\frac{d}{n},
\]
where $L_y(q_y )$ is ignored for model selection. Multiplying by $2n$ converts these approximations to
${\rm AIC}_{x;y}$ and AIC, respectively.

\section{PDIO and AICcd} \label{sec:pdio-aiccd}

The idea behind the derivation of PDIO and ${\rm AIC}_{cd}$ is to replace $\hat q_x$ by
\begin{equation} \label{eq:asm-hat-qx}
  \hat q_x = p_{z|y}(\bmh{\theta}_y) \hat q_y.
\end{equation}
This implies (\ref{eq:asm-qx}) by considering the limiting situation of $n\to\infty$.
Thus, the assumption for PDIO and ${\rm AIC}_{cd}$ is stronger
than the assumption for ${\rm AIC}_{x;y}$.
Substituting (\ref{eq:asm-hat-qx}) into the complete-data MLE gives
\begin{equation} \label{eq:mlex}
  \bmh{\theta}_x   = \argmin_{\bm{\theta}\in\Theta} L_x(\hat q_x ; p_x(\bm{\theta})).
\end{equation}
Comparing (\ref{eq:mlex}) with (\ref{eq:mley-em}) gives $\bmh{\theta}_x = \bmh{\theta}_y$. 
Therefore, there should not be any missing data, or at least $p_{z|y}(\bm{\theta})$ should not involve the parameter $\bm{\theta}$.
Consequently,  AIC, PDIO, ${\rm AIC}_{cd}$, and ${\rm AIC}_{x;y}$ are equivalent when
PDIO and ${\rm AIC}_{cd}$ are justified under (\ref{eq:asm-hat-qx}).

Although assumption (\ref{eq:asm-hat-qx}) is too strong to work with,
it is interesting to see how PDIO and ${\rm AIC}_{cd}$ would be derived if (\ref{eq:asm-hat-qx}) is formally accepted.
The argument below to derive PDIO and ${\rm AIC}_{cd}$ is rather confusing because $\hat q_x$ is interpreted interchangeably as the complete-data empirical distribution or the right-hand side of (\ref{eq:asm-hat-qx}).

By a similar argument to the proof of Theorem~\ref{thm:riskxy-hat}, the Taylor expansion of $L_x(\hat q_x ; p_x(\bm{\theta}))$ around $\bm{\theta}=\bmh{\theta}_y$ is
\begin{equation} \label{eq:lxx-taylor}
  L_x(\hat q_x ; p_x(\bm{\theta})) = 
  L_x(\hat q_x ; p_x(\bmh{\theta}_y)) +
  \frac{1}{2} (\bm{\theta} - \bmh{\theta}_y)^T \hat H_x  (\bm{\theta} - \bmh{\theta}_y)
  +O_p(n^{-3/2})
\end{equation}
with $\hat H_x = H_x(\hat q_x; \bmh{\theta}_y)$. Its expectation with $\bm{\theta}=\bmb{\theta}_y$ gives
\begin{equation} \label{eq:lxx-expand}
   L_x(q_x ; p_x(\bmb{\theta}_y)) 
   = E\{  L_x(\hat q_x ; p_x(\bmh{\theta}_y)) \} + \frac{1}{2n}\trace(H_x H_y^{-1} G_y H_y^{-1}) + O(n^{-3/2}).
\end{equation}
This corresponds to (\ref{eq:lyy-expand}) of Theorem~\ref{thm:riskxy-hat}.
Noticing  (\ref{eq:dxisdy}) and thus, $D_y(q_y ; p_y(\bmb{\theta}_y)) = L_x(q_x ; p_x(\bmb{\theta}_y)) - L_x(q_x)$, and then substituting (\ref{eq:lxx-expand}) into (\ref{eq:riskxy-expand}) gives the  estimator of  ${\rm risk}_{x;y}$ unbiased up to $O(n^{-1})$ under (\ref{eq:asm-hat-qx}) as
\begin{equation} \label{eq:aicxy-wrong}
    \widehat{\rm risk}_{x;y} =  L_x(\hat q_x ; p_x(\bmh{\theta}_y)  ) - L_x(q_x) 
     +\frac{1}{n} \trace(H_x H_y^{-1} G_y H_y^{-1}).
\end{equation}
The goodness of fit term is $ L_x(p_{z|y}(\bmh{\theta}_y) \hat q_y; p_x(\bmh{\theta}_y) ) =  - Q(\bmh{\theta}_y ; \bmh{\theta}_y) /n$ under (\ref{eq:asm-hat-qx}). Therefore, (\ref{eq:aicxy-wrong}) gives ${\rm AIC}_{cd}$ by the same approximation used to derive ${\rm AIC}_{x;y}$.

In \cite{cavanaugh1998akaike}, for evaluating (3.15) there, they assumed that $E\{ Q(\bm{\theta}_0 ;\bmh{\theta}_y ) \} \approx E\{ Q(\bm{\theta}_0 ;\bm{\theta}_0 ) \}$ or $E( L_x(p_{z|y}(\bmh{\theta}_y) \hat q_y; p_x(\bm{\theta}_0))) \approx L_x(p_{z|y}(\bm{\theta}_0) q_y; p_x(\bm{\theta}_0)) $ under the correct specification $q_x = p_x(\bm{\theta}_0)$.
The equality holds exactly under (\ref{eq:asm-hat-qx}) because $E(L_x(\hat q_x ; p_x(\bm{\theta}_0))) = L_x(q_x ; p_x(\bm{\theta}_0)) $ if $\hat q_x$ is interpreted as the empirical distribution.
Unfortunately, the difference is $E\{ Q(\bm{\theta}_0 ;\bmh{\theta}_y ) \} - E\{ Q(\bm{\theta}_0 ;\bm{\theta}_0 ) \} = O(1)$ in general without assuming (\ref{eq:asm-hat-qx}), leading to the bias of ${\rm AIC}_{cd}$ even when (\ref{eq:asm-qzy}) holds.

In \cite{shimodaira1994new}, (3.5) corresponds to our (\ref{eq:lxx-taylor}), where $\bmh{\theta}_x = \bmh{\theta}_y$ is assumed implicitly in order to ignore the first derivative.
Although $L_x(\hat q_x)$ diverges for continuous random variable $\bm{x}$,  $D_x(\hat q_x ; p_x(\bmh{\theta}_y)) =  L_x(\hat q_x ; p_x(\bmh{\theta}_y)  ) - L_x(\hat q_x)$ is formally considered.
Similar to (\ref{eq:dxisdy}), we then have $D_x(\hat q_x ; p_x(\bmh{\theta}_y))  = D_y(\hat q_y ; p_y(\bmh{\theta}_y))$ in (3.6) there.
From this argument, the goodness of fit term of (\ref{eq:aicxy-wrong}) is $ L_x(\hat q_x ; p_x(\bmh{\theta}_y)  ) = L_y(\hat q_y ; p_y(\bmh{\theta}_y)) +  L_x(\hat q_x) - L_y (\hat q_y)$, where $L_x(\hat q_x) - L_y (\hat q_y)$ is independent of the model specification if $\hat q_x$ is interpreted as the empirical distribution.
Therefore, (\ref{eq:aicxy-wrong}) gives PDIO because $L_x(\hat q_x ; p_x(\bmh{\theta}_y)  )$ can be replaced with $L_y(\hat q_y ; p_y(\bmh{\theta}_y)) $ for model selection.

\section{Simulation study} \label{sec:simulation}

\subsection{Simulation 1}

To verify Theorem~\ref{thm:riskxy-hat}, we performed a simulation study of the two-component normal mixture model defined as follows. Let $z\in \{1,2\}$ be
a discrete random variable for the component label, and $y\in \mathbb{R}$ be a continuous random
variable for the observation.  The distribution of $z$ is $P(z=i)=\pi_i$ and the
conditional distribution of $y$ given $z=i$ is the normal distribution with mean $\mu_i$ and
variance $\sigma_i^2$.
The true parameter for data generation is specified as $\bm{\theta}_0^T = (\pi_1, \mu_1, \mu_2,
\sigma_1^2, \sigma_2^2) = (0.6, -1, 1, 0.7^2, 0.7^2)$. We consider two candidate models for
selection. Model~1 is a two-component normal mixture model with a constraint $\sigma_1^2 =
\sigma_2^2$ ($d=4$), whereas Model~2 is the same model without the constraint ($d=5$). 
Because these two models are correctly specified, (\ref{eq:asm-qzy}) holds.
However, (\ref{eq:asm-hat-qx}) obviously does not.

We generated $B=4000$ datasets with sample size $n = 100, 200, 500, 1000, 2000, 5000, 10000$. They are denoted as
$X^{(b)}=(\bm{x}^{(b)}_1,\ldots, \bm{x}^{(b)}_n)$, $b=1,\ldots, B$. We also generated datasets of
sample size $\tilde{n}=15000$, which are denoted as
$\tilde{X}^{(b)}=(\bmt{x}^{(b)}_1,\ldots,\bmt{x}^{(b)}_{\tilde{n}})$ for computing the loss functions. 
For each $X^{(b)}=(Y^{(b)},Z^{(b)})$ and Model~$k$, $k=1,2$, we computed the information criteria ${\rm AIC}(Y^{(b)},
k)$, ${\rm PDIO}(Y^{(b)}, k)$, ${\rm AIC}_{cd}(Y^{(b)}, k)$, ${\rm AIC}_{x;y}(Y^{(b)}, k)$, and the
loss functions ${\rm loss}_{y;y}(Y^{(b)}, k)= L_y(q_y; p_y(\bmh{\theta}^{(b)}_y))$, ${\rm loss}_{x;y}(X^{(b)}, k)=L_x(q_x; p_x(\bmh{\theta}^{(b)}_y))$, where $\bmh{\theta}^{(b)}_y$ is computed from $Y^{(b)}$.
In the formulas below, $:\approx$ denotes that the expectation on the left-hand side is computed numerically by the simulation on the right-hand side.
The loss functions are computed numerically by
\begin{align*}
  {\rm loss}_{y;y}(Y^{(b)}, k) & 
  :\approx -\frac{1}{\tilde n} \sum_{t=1}^{\tilde n} \log p_y(\bmt{y}^{(b)}_t ; \bmh{\theta}^{(b)}_y),\\
  {\rm loss}_{x;y}(X^{(b)}, k) &
  :\approx -\frac{1}{\tilde n} \sum_{t=1}^{\tilde n} \log p_x(\bmt{x}^{(b)}_t ; \bmh{\theta}^{(b)}_y),
\end{align*}
where $p_x$, $p_y$, and $\bmh{\theta}^{(b)}_y$ are for Model~$k$.
Then the expectation with respect to $q_x=p_x(\bm{\theta}_0)$ is computed by the simulation average.
For example,
\begin{gather*}
  E(\Delta {\rm AIC}) :\approx \frac{1}{B} \sum_{b=1}^B
   ( {\rm AIC}(Y^{(b)},1)  - {\rm AIC}(Y^{(b)},2)   ), \\
     \Delta {\rm risk}_{x;y} :\approx \frac{1}{B} \sum_{b=1}^B
   ( {\rm loss}_{x;y}(X^{(b)},1) -  {\rm loss}_{x;y}(X^{(b)},2)    ).
\end{gather*}
This Monte Carlo method calculates the expectation accurately for sufficiently large $\tilde n$ and $B$.

The result shown in Table~\ref{tab:ex1} verifies Theorem~\ref{thm:riskxy-hat}.
For sufficiently large $n$, $E(\Delta
{\rm AIC}) = 2n \Delta {\rm risk}_{y;y}$ and $E(\Delta {\rm AIC}_{x;y}) = 2n \Delta {\rm
risk}_{x;y}$ hold very well.  On the other hand, 
$E(\Delta {\rm PDIO})$ differs significantly from $2n \Delta {\rm risk}_{y;y}$ and  $2n
\Delta {\rm risk}_{x;y}$. Thus, PDIO is not a good estimator of either of these risk functions.
In addition, the expected value of ${\rm AIC}_{cd}$ is similar to that of PDIO, but its
variation is larger than PDIO, as seen in the standard errors.

Let us consider the difference ${\rm PDIO} - {\rm AIC}_{cd}$
\[
  {\rm diff}(Y, \bmh{\theta}_y)  = 2Q(\bmh{\theta}_y; \bmh{\theta}_y) - 2\ell_y(\bmh{\theta}_y) = 
  2 \sum_{t=1}^n 
  \int p_{z|y} (\bm{z}| \bm{y}_t ; \bmh{\theta}_y)
  \log p_{z|y} (\bm{z}| \bm{y}_t ; \bmh{\theta}_y) \,d\bm{z},
\]
and its difference between the two models, which is denoted as $\Delta {\rm diff}(Y, \bmh{\theta}_y) = \Delta{\rm PDIO} - \Delta{\rm AIC}_{cd}$.
$\Delta {\rm diff}(Y, \bmh{\theta}_y)$ and $E(\Delta {\rm diff}(Y, \bmh{\theta}_y))$ can be very large, and they are $O(n)$ under model misspecification.
If (\ref{eq:asm-qzy}) holds, as is the case of Table~\ref{tab:ex1}, $E({\rm diff}(Y, \bmb{\theta}_y))=2n \int q_x(x) \log q_{z|y}(z|y)\,dx$ is independent of the model. Therefore, the difference becomes smaller; $\Delta {\rm diff}(Y, \bmh{\theta}_y)=O_p(\sqrt{n})$ and $E(\Delta {\rm diff}(Y, \bmh{\theta}_y))=O(1)$.

\def\arraystretch{1.1}
\begin{table}
  \caption{Expected values of the information criteria and the risk functions in Simulation~1.
   These values are differences between the two models with standard errors in parentheses.}
  \label{tab:ex1}
  \centering
  \begin{tabular}{cccccccc}
  \hline 
  $n$ & 100 & 200 & 500 & 1000
    & 2000 & 5000 & 10000 \\
  \hline 
  \multirow{1}{*}{$E(\Delta {\rm AIC})$} &
    0.810 & 0.898 & 0.982 & 0.978 & 
    0.986 & 0.982 & 1.04 \\ 
    & (0.027) & (0.025) & (0.023) & (0.023) & 
      (0.023) & (0.023) & (0.022) \\[0.5em]
%  \hline
  \multirow{1}{*}{$E(\Delta {\rm PDIO})$} &
    43.5 & 41.1 & 37.0 & 36.0 & 34.9 & 34.4 & 34.2 \\ 
    & (1.64) & (0.716) & (0.344) & (0.220) & 
      (0.141) & (0.088) & (0.064) \\[0.5em]
%  \hline
  \multirow{1}{*}{$E(\Delta {\rm AIC}_{cd})$} &
        42.3 & 41.0 & 37.2 & 36.6 & 35.2 & 35.5 & 33.5 \\ 
    & (1.67) & (0.793) & (0.518) & (0.494) & 
      (0.573) & (0.812) & (1.08) \\[0.5em]
%  \hline
  \multirow{1}{*}{$E(\Delta {\rm AIC}_{x;y})$} &
    22.1 & 21.0 & 19.0 & 18.5 & 
    18.0 & 17.7 & 17.6 \\ 
    & (0.821) & (0.361) & (0.174) & (0.113) & 
      (0.074) & (0.049) & (0.037) \\[0.5em]
  \hline
  \multirow{1}{*}{$2n \Delta {\rm risk}_{y;y}$} &
    1.83 & 1.47 & 1.15 & 1.08 & 
    1.03 & 1.02 & 0.967 \\ 
    & (0.052) & (0.040) & (0.030) & (0.027) & 
    (0.026) & (0.030) & (0.033) \\[0.5em]
%  \hline
  \multirow{1}{*}{$2n \Delta {\rm risk}_{x;y}$} &
    100.9 & 28.9 & 20.3 & 18.6 & 18.2 & 17.5 & 17.0 \\ 
    & (40.3) & (1.39) & (0.620) & (0.487) & 
    (0.456) & (0.464) & (0.430) \\ 
  \hline
  \end{tabular}
\end{table}

\subsection{Simulation 2}

We next performed a simulation study on the three-component normal mixture model to examine how well the information criteria work for model selection in a practical situation where some candidate models do not satisfy assumption (\ref{eq:asm-qzy}).
The true parameter value is
$\bm{\theta}_0^T = (\pi_1,\pi_2,\mu_1,\mu_2,\mu_3,\sigma_1^2,\sigma_2^2,\sigma_3^2)=
(0.5,0.3,-2,0,3,0.7^2,0.7^2,1^2)$. We consider five candidates with the following constraints.
Model~1 is $\sigma_1^2=\sigma_2^2=\sigma_3^2$ ($d=6$). Model~2 is $\sigma_2^2=\sigma_3^2$ ($d=7$).
Model~3 is $\sigma_1^2=\sigma_3^2$ ($d=7$). Model~4 is $\sigma_1^2=\sigma_2^2$ ($d=7$), and 
Model~5 has no constraint ($d=8$). 
Model~1, Model~2, and Model~3 are misspecified and do not satisfy (\ref{eq:asm-qzy}).
Model~4 and Model~5 are correctly specified and satisfy (\ref{eq:asm-qzy}).
None of the models satisfy (\ref{eq:asm-hat-qx}).
We have generated $B=10000$ datasets of $n=500$ and $\tilde n=2000$.

Table~\ref{tab:ex2_select} shows the model selection results. Model~4 is the best model in
the sense that it minimizes both ${\rm risk}_{y;y}$ and ${\rm risk}_{x;y}$
(Table~\ref{tab:ex2_risk}). All the information criteria tend to select Model~4.
AIC tends to choose a more complex model (i.e., Model~2 or Model~5) than the other criteria,
indicating a smaller penalty for model complexity. PDIO tends to choose a simpler model (i.e., Model~1),
implying a larger penalty for model complexity.

To compare candidate models in the long run, the expected loss of each Model~$k$ relative to that
of Model~4 is computed by
\[
   \Delta {\rm risk}_{x;y}(k) :\approx \frac{1}{B} \sum_{b=1}^B
   ( {\rm loss}_{x;y}(X^{(b)},k) -  {\rm loss}_{x;y}(X^{(b)},4)    ).
\]
Table~\ref{tab:ex2_risk} (upper) shows the results. The most complex model (Model~5) is the second
best in terms of ${\rm risk}_{y;y}$, but the simplest model (Model~1) is the second best in terms of
${\rm risk}_{x;y}$, indicating a large contribution of $p_{z|y}(\bm{\theta})$ to the second term of (\ref{eq:riskxy-expand}).

The information criterion performance is measured by the expected loss of the selected model.
For example, the performance of AIC in terms of complete data is measured by
\[
   \Delta {\rm risk}_{x;y}({\rm AIC}) :\approx \frac{1}{B} \sum_{b=1}^B
   ( {\rm loss}_{x;y}(X^{(b)},\hat k^{(b)}) -  {\rm loss}_{x;y}(X^{(b)},4)    ),
\]
where $\hat k^{(b)}$ is the minimum AIC model computed from $Y^{(b)}$.
Table~\ref{tab:ex2_risk} (lower) shows the results, where the value in bold denotes
the minimum value of each column. AIC outperforms the other criteria in terms of ${\rm
risk}_{y;y}$, and ${\rm AIC}_{x;y}$ outperforms the other criteria in terms of ${\rm risk}_{x;y}$.
In this example, some models do not satisfy assumption (\ref{eq:asm-qzy}), but AIC and ${\rm AIC}_{x;y}$ work very well as expected.

\begin{table}
  \caption{Frequency of model selection in Simulation~2.} \label{tab:ex2_select}
  \begin{threeparttable}
  \begin{tabular}{ccccccc}
  \hline
  & & Model~1 & Model~2 & Model~3 & Model~4\tnote{$*$} & Model~5\tnote{$*$} \\ 
  & & ($d=6$) & ($d=7$)  & ($d=7$)  & ($d=7$)  & ($d=8$)  \\ 
  \hline
  AIC  & & 881 & 2419 & 262 & 5600 & 838 \\
  PDIO & & 5442 & 16 & 4 & 4534 & 4  \\
  ${\rm AIC}_{cd}$ & & 2063 & 2 & 974 & 6551 & 410 \\ 
  ${\rm AIC}_{x;y}$ & & 3704 & 65 & 15 & 6190 & 26 \\
  \hline
  \end{tabular}
  \begin{tablenotes}
    \item[$*$] correctly specified model
  \end{tablenotes}
  \end{threeparttable} 
\end{table}

\begin{table}
    \caption{Risk functions for models and those for information criteria in Simulation~2.
    These values are relative to Model~4 with standard errors in parentheses.}
    \label{tab:ex2_risk} 
  \begin{threeparttable}
    \begin{tabular}{ccccccc} 
    \hline 
     & & \multicolumn{2}{c}{ $2n \Delta {\rm risk}_{y;y}$} & & 
         \multicolumn{2}{c}{ $2n \Delta {\rm risk}_{x;y}$} \\ 
    \hline 
    Model~1 & & 6.60 & (0.04) & & 33.2 & (0.21) \\ 
    Model~2 & & 1.40 & (0.02) & & 59.2 & (0.71) \\ 
    Model~3 & & 7.86 & (0.04) & & 80.7 & (0.80) \\ 
    Model~4\tnote{$*$} & &    0 & (0.00) & &    0 & (0.00) \\ 
    Model~5\tnote{$*$} & & 1.32 & (0.02) & & 45.6 & (0.87) \\ 
  \hline
    AIC               & & {\bf 1.44} & (0.03) & & 39.6 & (0.91) \\
    PDIO              & & 3.57 & (0.04) & & 19.6 & (0.30) \\
    ${\rm AIC}_{cd}$  & & 2.33 & (0.04) & & 28.2 & (0.72) \\ 
    ${\rm AIC}_{x;y}$ & & 2.36 & (0.04) & & {\bf 14.8} & (0.43) \\ 
    \hline 
    \end{tabular} 
  \begin{tablenotes}
    \item[$*$] correctly specified model
  \end{tablenotes}
  \end{threeparttable} 
\end{table}

\section{Concluding remarks} \label{sec:conclusion}

We derived ${\rm AIC}_{x;y}$ as an unbiased estimator of the expected
Kullback-Leibler divergence between the true distribution and the estimated distribution of
complete data when only incomplete data is available.
In Simulation~1, ${\rm
AIC}_{x;y}$  and AIC are unbiased up to the penalty terms, whereas PDIO and ${\rm AIC}_{cd}$ are not.

To derive ${\rm AIC}_{x;y}$, we assumed (\ref{eq:asm-qzy}), meaning that the conditional
distribution $p_{z|y}(\bm{\theta})$ of the missing data given the incomplete data is correctly
specified, while the marginal distribution $p_{y}(\bm{\theta})$ of the incomplete data is
misspecified in general. However, the conditional distribution is misspecified in practice. In
Simulation~2,  we observed that ${\rm AIC}_{x;y}$ and AIC perform better than the other criteria
even if some models are misspecified. Without assumption (\ref{eq:asm-qzy}), the dominant term in
(\ref{eq:riskxy-expand}) is $D_x(q_x; p_x(\bmb{\theta}_y))= D_x(q_x ; p_{z|y}(\bmb{\theta}_y) q_y)
+ D_y(q_y ; p_y(\bmb{\theta}_y)  )  \ge D_y(q_y ; p_y(\bmb{\theta}_y) )$. Thus, ${\rm
AIC}_{x;y}$ estimates the lower bound of $2n\, {\rm risk}_{x;y}$. 
It is impossible to reasonably estimate the ignored term $ D_x(q_x ; p_{z|y}(\bmb{\theta}_y) q_y)$ in our setting where $z_1,\ldots, z_n$ are missing completely.

Although we assume that $p_{z|y}(\bm{\theta})$ is correctly specified, it is beneficial to include $p_{z|y}(\bm{\theta})$ as a part of $p_x(\bm{\theta}) = p_{z|y}(\bm{\theta}) p_y(\bm{\theta})$ for model selection.
The variance of $\bmh{\theta}_y$ causes $p_{z|y}(\bmh{\theta}_y)$ to fluctuate even if $p_{z|y}(\bmb{\theta}_y)=q_{z|y}$.
The amount of this random variation is measured by the additional penalty term (\ref{eq:aicxy-aic}) in ${\rm AIC}_{x;y}$.

In the future, we plan to work on more complicated missing
mechanisms or combine a missing mechanism with other sampling mechanisms, such as the
covariate-shift \citep{shimodaira2000improving} problem. One important extension is 
semi-supervised learning \citep{chapelle2006semi,kawakita2014safe}, where the log-likelihood
function is
\[
  \ell(\bm{\theta}) =
   \sum_{t=1}^n \log p_y(\bm{y}_t; \bm{\theta}) + \sum_{t={n+1}}^{n+n'}
    \log p_x(\bm{x}_t; \bm{\theta}).
\]
In this case, the additional complete data $\bm{x}_{n+1},\ldots,\bm{x}_{n+n'}$ helps estimate
conditional distribution $q_{z|y}$. We may reasonably estimate $ D_x(q_x ; p_{z|y}(\bmb{\theta}_y) q_y)$ without assuming (\ref{eq:asm-qzy}), leading to a new information criterion, which 
will be the subject in future research.

\section*{Acknowledgments}

We would like to thank the reviewers for their comments to improve the manuscript.
We appreciate Kei Hirose and Shinpei Imori for their suggestions and comments. While preparing an earlier
version of the manuscript, which was published as \cite{shimodaira1994new}, Hidetoshi Shimodaira is indebted
to Shun-ichi Amari for the geometrical view of the EM algorithm and to Noboru Murata for the
derivation of the Takeuchi information criterion.

\appendix

\section{Technical details} \label{sec:technical}

\subsection{Proof of Lemma~\ref{lem:dx-decomp}} 
For brevity, we omit $(\bm{y},\bm{z})$ of $f_x(\bm{y},\bm{z})$ in the integrals below.
$D_x(g_x;f_x) = \int \int g_{z|y} g_y ( \log g_{z|y} + \log g_y - \log f_{z|y} - \log f_y
)d\bm{z}d\bm{y} = \int  g_y \int g_{z|y} ( \log g_{z|y} - \log f_{z|y} )d\bm{z}d\bm{y} + \int  g_y
(\int g_{z|y}d\bm{z}) ( \log g_y - \log f_y )d\bm{y} = \int  g_y \int g_{z|y} ( \log g_{z|y} g_y -
\log f_{z|y} g_y )d\bm{z}d\bm{y} + \int  g_y ( \log g_y - \log f_y )d\bm{y} = D_x(g_{z|y}g_y;
f_{z|y}g_y) + D_y(g_y; f_y)$, thus showing (\ref{eq:dx-decomp1}). $D_y(g_y; f_y) =  \int \int
h_{z|y} g_y (\log g_y - \log f_y + \log h_{z|y} -  \log h_{z|y}) d\bm{z} d\bm{y} = D_x(h_{z|y}g_y ;
h_{z|y} f_y)$, which shows (\ref{eq:dy-dx}).

\subsection{Proof of Lemma~\ref{lem:riskxy}} 
We assume $q_{z|y}=p_{z|y}(\bmb{\theta}_y)$ and $\bmb{\theta}_x = \bmb{\theta}_y$.
From the definitions of $\bmb{\theta}_x$ and $H_x$, we have
\[
  \frac{\partial D_x(q_x; p_x(\bm{\theta}))}{\partial \bm{\theta} }\Bigr|_{\bmb{\theta}_y} =0,
  \quad
  \frac{\partial^2 D_x(q_x; p_x(\bm{\theta}))}
  {\partial \bm{\theta} \partial \bm{\theta}^T}\Bigr|_{\bmb{\theta}_y} = H_x.
\]
Hence, the Taylor expansion of $D_x(q_x; p_x(\bm{\theta}))$ around $\bm{\theta}=\bmb{\theta}_y$ is
\[
  D_x(q_x; p_x(\bm{\theta}) )= D_x(q_x; p_x(\bmb{\theta}_y))
   + \frac{1}{2} (\bm{\theta} - \bmb{\theta}_y)^T
       H_x  (\bm{\theta} - \bmb{\theta}_y) + O(n^{-3/2})
\]
for $\bm{\theta} - \bmb{\theta}_y = O(n^{-1/2})$. The first
term on the right-hand side is $ D_y(q_y ; p_y(\bmb{\theta}_y)  )$ as shown in (\ref{eq:dxisdy}). 
Substituting $\bm{\theta} =
\bmh{\theta}_y$ in $ D_x(q_x; p_x(\bm{\theta}) )$ and taking its expectation
gives (\ref{eq:riskxy-expand}) by noting
\[
  E \left\{ (\bmh{\theta}_y - \bmb{\theta}_y)^T  H_x
   (\bmh{\theta}_y - \bmb{\theta}_y)  \right\}
  = \trace \left(  H_x   \, E \left\{ 
   (\bmh{\theta}_y - \bmb{\theta}_y)  (\bmh{\theta}_y - \bmb{\theta}_y)^T \right\} \right),
\]
which becomes $\trace  \left(   H_x H_y^{-1} G_y H_y^{-1}    \right) /n+ O(n^{-2})$ from
(\ref{eq:mley-normal}).

\subsection{Proof of Theorem~\ref{thm:riskxy-hat}} 

From the definitions of $\bmh{\theta}_y$ and $\hat H_y = H_y(\hat q_y; \bmh{\theta}_y)$, we have
\[
  \frac{\partial L_y(\hat q_y; p_y(\bm{\theta}))}
  {\partial \bm{\theta} }\Bigr|_{\bmh{\theta}_y} =0, \quad
  \frac{\partial^2 L_y(\hat q_y; p_y(\bm{\theta}))}
  {\partial \bm{\theta} \partial \bm{\theta}^T}\Bigr|_{\bmh{\theta}_y} = \hat H_y.
\]
Hence, the Taylor expansion of $L_y(\hat q_y ; p_y(\bm{\theta}))$ around
$\bm{\theta}=\bmh{\theta}_y$ is
\[
  L_y(\hat q_y ; p_y(\bm{\theta})) = 
  L_y(\hat q_y ; p_y(\bmh{\theta}_y)) +
  \frac{1}{2} (\bm{\theta} - \bmh{\theta}_y)^T \hat H_y  (\bm{\theta} - \bmh{\theta}_y)
  +O_p(n^{-3/2})
\]
for $\bm{\theta} - \bmh{\theta}_y = O_p(n^{-1/2})$. Substituting
$\bm{\theta}=\bmb{\theta}_y$ in $L_y(\hat q_y ; p_y(\bm{\theta}))$, we take its expectation below.
By noting $\hat H_y = H_y + O_p(n^{-1/2})$, we have
\[
  E \left\{ (\bmb{\theta}_y - \bmh{\theta}_y)^T \hat H_y 
     (\bmb{\theta}_y - \bmh{\theta}_y) \right\} = 
  \trace \left(  H_y \, E \left\{  (\bmh{\theta}_y - \bmb{\theta}_y)
     (\bmh{\theta}_y - \bmb{\theta}_y)^T \right\}  \right)  + O(n^{-3/2}),
\]
which becomes $ \trace(H_y H_y^{-1} G_y H_y^{-1} )/n + O(n^{-3/2}) $ from (\ref{eq:mley-normal}).
This proves (\ref{eq:lyy-expand}) because
\[
  E\{ L_y(\hat q_y ; p_y(\bmb{\theta}_y)) \}
  = E\{  L_y(\hat q_y ; p_y(\bmh{\theta}_y)) \} + \frac{1}{2n}\trace( G_y H_y^{-1}) + O(n^{-3/2}),
\]
and $E\{ L_y(\hat q_y ; p_y(\bmb{\theta}_y)) \} = L_y(q_y ; p_y(\bmb{\theta}_y))$. 
Substituting (\ref{eq:lyy-expand}) into (\ref{eq:riskxy-expand}) and comparing it
with (\ref{eq:hat-riskxy}) yields (\ref{eq:riskxy-unbiased}).

\bibliographystyle{imsart-nameyear}
\bibliography{stat2015}

\end{document}